\documentclass[aps,prb,twocolumn,showpacs,preprintnumbers,amsmath,amssymb,superscriptaddress]{revtex4}
\usepackage{color}
\usepackage{graphicx}
\usepackage{dcolumn}
\usepackage{bm}

\usepackage[hypertex]{hyperref}
\newcommand{\nc}{\newcommand}
\nc{\be}{\begin{eqnarray}}
\nc{\ee}{\end{eqnarray}}
\nc{\bea}{\begin{eqnarray}}
\nc{\eea}{\end{eqnarray}}
\nc{\bean}{\begin{eqnarray*}}
\nc{\eean}{\end{eqnarray*}}
\nc{\mb}{\mbox}
\nc{\rnc}{\renewcommand}
\nc{\vk}{\mb{\bf k}}
\nc{\vp}{\mb{\boldmath$p$}}
\nc{\rr}{\mb{\boldmath$r$}}
\nc{\vR}{\mb{\boldmath$R$}}
\nc{\vz}{\hat {\mb{\bf z}}}
\nc{\vj}{\mb{\boldmath$j$}}
\nc{\vg}{\mb{\boldmath$g$}}
\nc{\vE}{\mb{\boldmath$E$}}
\nc{\vB}{\mb{\boldmath$B$}}
\nc{\vH}{\mb{\boldmath$H$}}
\nc{\vM}{\mb{\boldmath$M$}}
\nc{\vP}{\mb{\boldmath$P$}}
\nc{\vS}{\mb{\boldmath$S$}}
\nc{\x}{\mb{\boldmath$x$}}
\nc{\A}{\mb{\boldmath$A$}}
\nc{\va}{\mb{\boldmath$a$}}
\nc{\vq}{\mb{\boldmath$q$}}
\nc{\vn}{\mb{\boldmath$n$}}
\nc{\vs}{\mb{\boldmath$\sigma$}}
\nc{\vt}{\mb{\boldmath$\tau$}}
\nc{\vpi}{\mb{\boldmath$\pi$}}
\nc{\nab}{\bm{\nabla}}
\nc{\X}{\sf x}

\begin{document}

\title{
Surface Quantized Anomalous Hall Current and Magneto-Electric Effect\\ 
in Magnetically Disordered Topological Insulators
}

\author{Kentaro Nomura}
\affiliation{
Correlated Electron Research Group (CERG), RIKEN-ASI, Wako 351-0198, Japan
            }
\author{Naoto Nagaosa}
\affiliation{
Correlated Electron Research Group (CERG), RIKEN-ASI, Wako 351-0198, Japan
            }
\affiliation{
Cross-Correlated Material Research Group (CMRG), RIKEN-ASI, Wako 351-0198, Japan
}
\affiliation{Department of Applied Physics, The University of Tokyo, Hongo, Bunkyo-ku, Tokyo 113-8656, Japan}

\date{\today}

\begin{abstract}
We study theoretically the role of quenched magnetic disorder at the surface of
a topological insulator by numerical simulation and scaling analysis.
It is found that all the surface states are localized while the 
transverse conductivity is quantized to be $\pm { {e^2} \over {2 h}}$
as long as the Fermi energy is within the bulk gap.
This greatly facilitates the realization of the topological 
magnetoelectric effect proposed by Qi {\it et al.} 
(Phys. Rev. B\textbf{78}, 195424 (2008)) with the surface magnetization 
direction being controlled by the simultaneous application of magnetic 
and electric fields .
\end{abstract}

\maketitle


The bulk-surface correspondence has an essential role in a large variety of phenomena in condensed matter physics, such as ferroelectricity, diamagnetism, the Meissner effect, and the quantum Hall effect.
The topological magnetoelectric (ME) effect is a novel manifestation of the bulk-surface correspondence in which the bulk magnetization is generated by a circulating quantized Hall current flowing at the surface of topologically nontrivial insulators, called topological insulators (TIs)\cite{review_TI}.

The electromagnetic response of a three-dimensional (3D) TI 
can be described by the Lagrangian for the axion electrodynamics, \cite{review_TI,Qi_2008,Essin_2009,PALee_2010}
\bea
 {\cal L}= \frac{1}{8\pi} 
\Big(\epsilon \vE^2-\frac{1}{\mu}\vB^2\Big)
+
\Big(\frac{\alpha}{4\pi^2}\Big)\,\theta\, \vE\!\cdot\!\vB,
\label{eq:action}
\eea
where $\vE$ and $\vB$ are the electromagnetic fields, 
$\epsilon$ and $\mu$ are the dielectric constant and magnetic permeability,
and $\alpha=e^2/\hbar c$ is the fine structure constant.
The first term is the conventional Maxwell term. 
The second term, called the $\theta$-term\cite{Wilczek_1987}, 
characterizes the topological nature of three-dimensional insulators; $\theta=0$ or $\pi$ (mod $2\pi$) corresponding to ordinary insulators and topological insulators, respectively.
The $\theta$ term is written as the total divegence,
and hence can be transformed into the surface Chern-Simons term in a system terminated by a boundary.
When a topological insulator is surrounded by a ferromagnetic layer, time-reversal ($T$) symmetry is broken, and the value of $\theta$ (more precisely $\nab\theta$) can be determined as discussed in Refs. \onlinecite{Qi_2008,Essin_2009}. 
A circulating Hall current, induced by an applied electric field\cite{Wilczek_1987,Jackiw_1984} on the surface, is the source of a bulk magnetization\cite{Qi_2008}.
These features are distinct from the
conventional ME effect\cite{Essin_2009} which is a long-term issue in the field of multiferroics.

\begin{figure}[b]
\begin{center}
\includegraphics[width=0.4\textwidth]{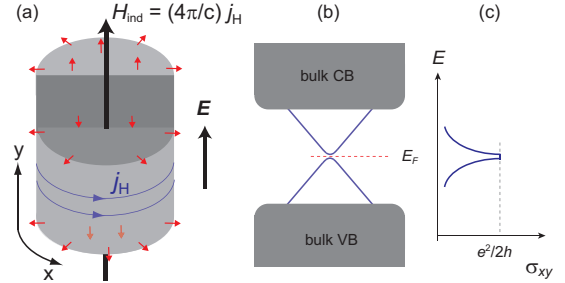}
\caption{(color online)
(a) Illustration of a magnetically doped topological 
insulator with a cylindrical geometry.
(b) Surface massive Dirac dispersion. The Hall 
conductivity $\sigma_{xy}$ is quantized when the 
Fermi level lies in the surface gap. $\sigma_{xy}$ 
deviates from $e^2/2h$ when the Fermi level is out of the surface gap.
}
\label{F1}
\end{center}
\end{figure}

Experimentally, however, there are several difficulties to realize
this topological ME effect. (a) First, it is required to get rid of the bulk
carriers. (b) Second, one needs to attach the insulating ferromagnetic layer
with the magnetization normal to the surface all pointing out
or in. (c) Lastly, the Fermi energy must be tuned accurately 
within the small gap of the surface Dirac fermion
opened by the exchange interaction. Otherwise, the description by 
eq.(\ref{eq:action}) is not justified. 
(a) is progressively realized\cite{Ando_2010}, while (b) and (c) seem still very difficult
at the moment, even though a recent experimental work
shows that the gapless surface Dirac 
states of the pristine topological insulator Bi$_2$Se$_3$ become 
gapped upon introducing magnetic impurities (Mn and Fe) 
into the crystal and also Mn dopants lead carrier doping
\cite{Chen_2010}. 
Therefore, the topological ME appears to be not practical
even though several theoretical proposals have been made
\cite{Qi_2008,Liu_2009,
Guo_2010,Garate_2010,Nomura_2010,Nunez_2010,Abanin_2010,Tse_2010,Maciejko_2010}.

In this paper, we study the effects of the quenched magnetic impurities or disorder 
on the surface of TI. In sharp contrast to conventional quantum Hall 
systems, all the surface states are localized while the 
Hall conductivity is quantized to be  
$\pm { {e^2} \over {2 h}}$
as long as the Fermi energy is within the bulk gap.
This resolves the problem (c). 
Consequently the generated magnetization is robust over randomness and universal.
Furthermore, it is shown that
this also resolves (b) with the simultaneous application 
of magnetic and electric fields parallel or antiparallel 
to each other. By this method, the surface magnetization 
can be controlled by {\it the bulk energy}, and hence
can easily overcome the magnetic anisotropy and Zeeman splitting {\it at the surface}.

\begin{figure}[b]
\begin{center}
\includegraphics[width=0.45\textwidth]{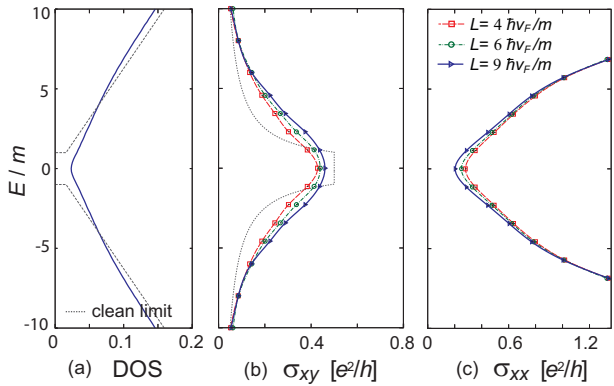}
\caption{(color online)
(a) Density of states, (b) Hall conductivity, and (c) diagonal conductivity of massive Dirac fermions are shown as a function of the Fermi energy divided by the mass gap $E/m$. Lines are guides for the eyes.
System sizes are $L=4\hbar v_F/m,6\hbar v_F/m$, and $9\hbar v_F/m$
}
\label{F2}
\end{center}
\end{figure}

\begin{figure}[b]
\begin{center}
\includegraphics[width=0.37\textwidth]{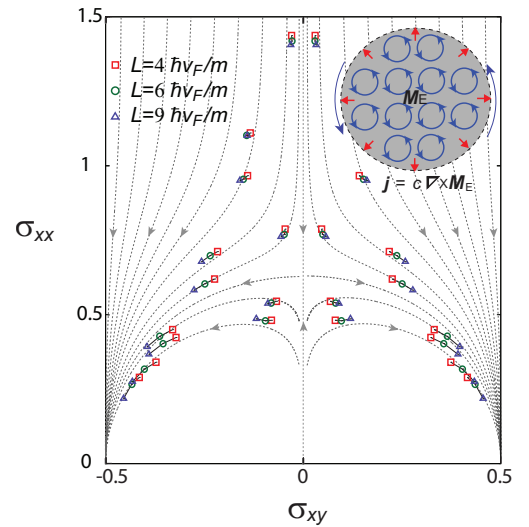}
\caption{(color online)
Scaling flow of $\sigma_{xx}$ and $\sigma_{xy}$ as increasing system size, indicating $(\sigma_{xx},\sigma_{xy})\rightarrow(0,\pm 1/2)$, in units of $e^2/h$, at $L\rightarrow\infty$. Curves are guides for the eyes.
Inset: The surface Hall current is related to the bulk orbital magnetization induced by an electric field.
}
\label{F3}
\end{center}
\end{figure}

We study the surface transport of a 3D TI 
basing on the 2D massive Dirac model with various types of disorder.
The interaction between magnetic dopants and surface electrons is 
described by the exchange Hamiltonian:\cite{Liu_2009,Guo_2010,Nunez_2010,Abanin_2010,Yang_2010}
$
 {\cal H}_{\rm exc}=-J\sum_{i=1}^{N_{\rm imp}}\vS_i \cdot \vs \delta(\rr-\vR_i).
$
The uniform part of $z$-component of local spins generates a mass gap 
$m^{}\equiv J n_{\rm imp}\overline{S}_z$ in the surface spectrum\cite{Qi_2008}, 
where $\vs=(\sigma_1^{},\sigma_2^{},\sigma_3^{})$ are Pauli matrices that 
act on the electron spin degrees of freedom, $\vS_i$ and $\vR_i$ are the local spin 
operator and the position operator of $i$th magnetic dopant, $N_{\rm imp}$ and $n_{\rm imp}$ are the 
total number and the mean sheet density of local spins.
The surface Dirac modes can be described by the Dirac Hamiltonian:\cite{review_TI}
\bea
 {\cal H}_{\rm Dirac}^{2D}=-i\hbar v_F\vz\times\vs\cdot\nab+m\sigma^{}_3.
\eea
On the other hand, 
 inhomogeneous part of the local magnetization gives disorder term:
${\cal V}=\sum_{\mu=0}^3 \sigma_{\mu} V_{\mu}(\rr)$, where $V_1$ and $V_2$ have a role of random vector potential, and $V_3$ has a role of random mass potential. 
The scalar potential $V_0(\rr)$ also could be introduced by impurities or vacancies. 
Here $\sigma_0$ is a $2\times2$ unit matrix.

In the clean limit, the Hall conductivity is quantized as
$ \sigma_{xy}=-{\rm sgn}(m)\frac{e^2}{2h}$
when the Fermi level is in the middle of the surface gap ($|E| <|m|$)~\cite{review_TI,Jackiw_1984,Niemi_1983,AHE}. 
The fact that $\sigma_{xy}$ is quantized even in the limit of 
$m\rightarrow 0$ is referred to parity anomaly\cite{Niemi_1983}.
When $|E|>|m|$, on the other hand, $\sigma_{xy}$ deviates from the quantized value, and vanishes at $|E|/|m| \rightarrow\infty$ as sketched in the Fig.1(c).\cite{AHE}

We evaluate the diagonal and the Hall conductivity of random massive Dirac Hamiltonian with the Kubo formula\cite{AHE}
\bea
 \sigma_{ab}(L)=-\frac{i\hbar}{L^2}\sum_{n,n'}\frac{f(E_n)\!-\!\!f(E_{n'})}{E_n-E_{n'}}
\frac{\langle n|j_{a}|n'\rangle\langle n'|j_{b}|n\rangle}{E_n-E_{n'}+i\eta}
\eea
where $a, b=x$ or $y$, $j_{a}$  is the current operator, and $|n\rangle$ denotes an eigenstate with its eigenvalue $E_n$ of Hamiltonian ${\cal H}^{2D}_{\rm Dirac}+{\cal V}$.
We work in the momentum space by introducing a hard cutoff at a sufficiently large momentum $\Lambda$.\cite{Nomura_2007,Nomura_2008}
Eigenstates and eigenvalues are obtained by numerically diagonalizing the Dirac Hamiltonian with disorder terms in the momentum space.
Random averaging is taken over typically 1000-10000 disorder configurations.
We use the Gaussian model for disorder potentials which obey $\langle V_{\mu}(\vq) V_{\nu}(\vq')\rangle=\delta_{\mu\nu}g_{\mu}\exp(-\vq^2d^2/2)\delta(\vq+\vq')$ with $\mu,\nu=0,1,2,3$. We set $g_0=g_1=g_2=g_3$ as assumed in realistic situations\cite{Chen_2010}
 and the disorder strength so that the disorder broadening energy is the order of the surface gap $m$ as seen in the density of states shown in Fig. 2(a).

The diagonal $\sigma_{xx}$ and the Hall conductivity $\sigma_{xy}$ are shown in Fig.2 for three different system sizes as a function of $E/m$, $E$ being the Fermi energy. 
The sign of $\sigma_{xy}$ is determined solely by the sign of the mass, no matter whether the Fermi level resides in the electron- or hole-region.
The size dependence indicates that  $\sigma_{xx}$ decreases while $\sigma_{xy}$ increases as the size $L$ increases.
The linear size of the system is characterized by the momentum cutoff $\Lambda$ or the mass gap $m$. Three system sizes shown in Fig.2 are $L=75\Lambda^{-1}= 4\hbar v_F^{}/m$, $113\Lambda^{-1}=6\hbar v_F^{}/m$, and $176\Lambda^{-1}=9\hbar v_F^{}/m$.

To see the tendency of $\sigma_{xy}$ and $\sigma_{xx}$ in the $L\rightarrow\infty$ limit, we plot $(\sigma_{xx},\sigma_{xy})$ by changing the system size $L$ in Fig. 3.
Although the system size cannot be widely changed because of the computational limitation,
 Fig.3 indicates that, with increasing system size, $(\sigma_{xx},\sigma_{xy})$ flows and approaches to two fixed points $(0,\pm 1/2)$ in units of $e^2/h$.
The scaling law of $\sigma_{xx}$ and $\sigma_{xy}$ has been originally studied in the integer quantum Hall effect\cite{QHE,QHE2,Dolan_1999} with corresponding fixed points $(0,n)$ in units of $e^2/h$, $n$ being an integer. 
In the present case with single-flavor Dirac fermions, a field theoretical study\cite{Ostrovsky_2007} shows that the values of $\sigma_{xy}$ shifts by 1/2 from the conventional ones, consistent with Fig.3.
The 1/2-shift results in a marked point, that is, $\sigma_{xy}=0$ is unstable.
This means that even if the surface mass gap is, in principle, infinitely small, $|\sigma_{xy}|$ increases and approaches to $e^2/2h$. This is a generalization of parity anomaly to the case with disorder. This is in contrast to the multi-Dirac-cone 2D lattice systems where mixing between different Dirac-cones wipes out the Hall conductivity quantization.\cite{Ostrovsky_2007}

When the system size exceeds the localization length, the localization effect becomes important and $\sigma_{xy}$ starts to flow toward the two fixed points as shown above.
The maximum system size in above analysis corresponds to $\sim 0.2[\mu {\rm m}]$ (with identifying $\Lambda$ with the largest wave number in the surface spectrum $\sim 0.2 [\AA ^{-1}]$)  which is very small compared to the realistic sample size and also to the coherence length. 
We expect that in the experimental situations the quantization of the Hall conductivity should be more prominent. In conventional quantum Hall systems, recent experiments\cite{Murzin_2005}
explored the temperature driven flow diagram of $\sigma_{xx}$ and $\sigma_{xy}$ in a large-temperature range from 4 K down to 40 mK. Similar scaling properties, but different fixed points, are expected on a magnetically doped surface of a topological insulator.

We note that even for states out of the original surface gap, which were metallic in the clean limit, $(\sigma_{xx},\sigma_{xy})$ scales to the fixed points $(0,\pm 1/2)$.
 This means the fact that all the surface states are localized, while a transverse current flows.
 It has been known, however, that in two-dimensional electron gas systems at least one extended state below the Fermi level is required to realize the quantum Hall effect\cite{QHE}, otherwise $\sigma_{xy}$ vanishes. 
Since all states of disordered massive Dirac Hamiltonian are localized in zero magnetic field, a finite quantized Hall transport sounds enigmatic: what carries the electric charges?
In the following, we address this problem basing on the bulk-surface correspondence.

Since all the surface states are localized by magnetic disorder, the
description by the $\theta$-term in eq.(\ref{eq:action}) is justified,
which corresponds to the magnetoelectric effect:
\bea
 \vM_{\!E}^{}=\Big(\frac{\alpha}{4\pi^2}\Big)\theta\vE, \qquad
 \vP_{\!B}^{}=\Big(\frac{\alpha}{4\pi^2}\Big)\theta\vB.
\label{ME}
\eea
These results have been reproduced from the explicit Wannier state representation of a topological insulator.\cite{Malashevich_2010,Essin_2010,Coh_2010}
Here $\theta$ is only well defined as a bulk property modulo $2\pi$.
The integer multiple of $2\pi$ can be specified once we specify a particular way to make the boundary with broken $T$-symmetry. More appropriately, the current and charge densities are given by 
\bea
\vj=c\nab\times\vM_{\!E}+\partial\vP_{\!B}/\partial t,\quad \rho=-\nab\cdot\vP_{\!B},
\label{j}
\eea
consistent with the half-integer quantized Hall surface states, under the condition\cite{Qi_2008,Wilczek_1987,Tse_2010,Essin_2010} $d\theta/dz=\sigma_{xy}(2\pi h/e^2)\delta(z-R)$, where $z=R$ is at the surface shown in Fig.1(a) and the inset in Fig.3.
Since bulk moments Eqs.(\ref{ME}) are protected by the bulk gap (when it remains intact) and insensitive to disorder on the surface, Eqs.(\ref{j}) can survive even though all surface states are localized.
With above results, the existence of the surface quantum Hall states, generalizes the axion electrodynamics effects to disordered systems.
Precisely, the energy region for the topological ME effect, which was originally limited in the surface gap, is enlarged to the bulk gap by localization effect of surface states. 
This will greatly facilitate the experimental realization.


Even though the 2D surface Dirac Hamiltonian omits such the bulk properties by the construction, non-vanishing half-quantized Hall conductivity is represented as parity anomaly.
A similar situation can be seen in the context of chiral anomaly. 
For the one-dimensional edge channel of the quantum Hall system, the chiral anomaly represents the current flow from or to 2D bulk region of the sample.
\cite{Nagaosa_1995}

\begin{figure}[b]
\begin{center}
\includegraphics[width=0.35\textwidth]{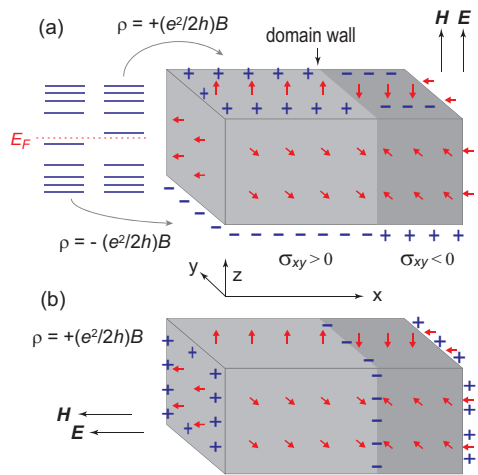}
\caption{(color online)
 Illustration of domain structures and electromagnetic control of them.
Domain wall is charged in the case of (b).
}
\label{F4}
\end{center}
\end{figure}

We note that the topological ME effect requires a finite surface gap with an entirely definite sign of the mass, otherwise there exists a massless one-dimensional channel at the domain boundary separating different signs of the mass.\cite{review_TI,Qi_2008}
 In general, however, magnetic moments of impurities could form a domain structure when temperature decreases down below the transition temperature\cite{Liu_2009,Guo_2010,Garate_2010,Nomura_2010,Nunez_2010,Abanin_2010}.
Simple situations are sketched in Fig.4. 
 There are two (many in general) domain regions where the surface magnetizations point alternative directions, one $S_{\perp}>0$ (negative mass) and the other $S_{\perp}<0$ (positive mass). 

In the remaining part of the paper, we propose a practical way to control the domain structure using the massive Dirac surface states.
First we assume that a magnetic field is applied in $+z$ direction as shown in Fig.4(a). For simplicity, let us start with the pure case without disorder.
On the surfaces perpendicular to the magnetic field, the Landau level structure is formed.
 When the Fermi energy resides in the mass gap ($|E|<|m|$), the electric charge density $\rho=\pm(e^2/2h)B$ is generated on the top and bottom. The sign of the charge depends on the sign of the mass and also the direction of the magnetic field.
By applying an electric field in $z$ direction, the degeneracy between the right and left regions is lifted, because of the electric polarization energy. 
In the positive mass region there is an energy gain, while loss in the negative region, as estimated as
\bea
 U_{\theta}&=&-\int d^3x \Big(\frac{\alpha}{4\pi}\Big)\vE\cdot\vB \nonumber \\
&\simeq&
-10^{10}\big(E[{\rm V/cm}]\big)\big(B[{\rm T}]\big)
\big(L[{\rm cm}]\big)^3
\ \ [{\rm eV}],
\eea
where $\vB=\vH+4\pi{\overline {\vM}_{\rm imp}}$ includes the contribution from the bulk magnetization of impurities ${\overline {\vM}_{\rm imp}}$, and $L$ is the linear size of the domain region.
For this energy gain, surface electrons in the right region are transferred from top to bottom.
This process requires larger $U_{\theta}$ than the anisotropic energy\cite{Nunez_2010}
$
 U_{\rm aniso}/L^2\sim 5\times 10^{10}\ \  [{\rm eV/cm^2}]
$
and the Zeeman energy $U_{\rm Zeeman}/L^2\sim10^{9}(B[{\rm T}])\ \  [{\rm eV/cm^2}]$ to flip the surface magnetization (sign change of the mass) 
in the right region.  
Typical strengths of electric $E\sim 10^3$[V/cm] and magnetic fields $B\sim 1$ [T] are enough to flip and rearrange the surface magnetization.
 In the case of strong disorder so that the Landau level spacing is dominated by disorder broadening energy,
 the extended states originally located at the center of each Landau level levitate to high (low) energy regime if its energy in the clean limit was positive (negative). \cite{Note1}
Eventually, only localized states remain in the spectrum where $\sigma_{xy}$ scales to $\pm e^2/2h$ as shown above.
In this case, the induced charge density on the surface is $\rho=\sigma_{xy}B$ independent of the Fermi level as long as it is within the bulk gap.
Consequently, as domain wall moves, the positive mass region dominates over the negative mass region, and the uniform surface gap with a definite sign is attained, where the Hall current can flow circularly.
Once the uniform mass rearrangement is formed, the external fields $\vE$ and $\vH$ can be gradually turned off.
When an electric and magnetic fields point in $x$ direction (see Fig.4(b)), a similar effect occurs. The surfaces with generated electric charge are different but the direction in which domain walls move is same as the case of Fig.4(a).

In this work, we studied the effects of quenched magnetic disorder on the surface of a topological insulator, basing the surface Dirac model. 
The scaling analysis indicates that
 all surface states are localized, while the Hall conductivity approaches to the quantized value, even when the mass gap is smeared by disorder broadening. Consequently the plateau width at large system size and at low temperature is enhanced from that in the clean limit. 
This helps the experimental realization of the topological ME effect in a magnetically doped TI. We also proposed the way to control domain structures of the surface magnetization by the simultaneous application of electric and magnetic fields. 
The effective theory of TI including the magnetic impurities is an interesting and important issue, which is left as a future subject.

We are grateful to A. Furusaki, X. F. Jin, Q. Niu, S. Ryu, and D. Vanderbilt for useful discussion.
This work is supported by MEXT Grand-in-Aid No.20740167, 19048008, 19048015, 21244053, Strategic International Cooperative Program (Joint Research Type) from Japan Science and Technology Agency, and by the Japan Society for the Promotion of Science (JSPS) through its ``Funding Program for World-Leading Innovative R\&D on Science and Technology (FIRST Program)''.

\end{document}